Reduction in respiratory mortality not associated with influenza in Russia: effect of the introduction of pneumococcal vaccination (PCV13), or improvement in care?

Edward Goldstein[1,*]

1. Center for Communicable Disease Dynamics, Department of Epidemiology, Harvard TH Chan School of Public Health, Boston, MA 02115 USA
*. egoldste@hsph.harvard.edu

**Abstract**

*Background.* Pneumococcal vaccination (PCV13) for children (as well as older adults) in Russia was introduced in 2014, with no prior PCV7 use. While pneumonia hospitalization rates, both in children and adults didn't decrease in the following years, respiratory mortality rates did decline with time. Moreover, there is a strong association between antibiotic, including multidrug resistance and PCV13 vaccine serotypes for *S. pneumoniae* in children in Russia, and presence of *S. pneumoniae* among sepsis cases in Russia during the recent years has been very low. Annual variability in influenza circulation may affect some of the changes in respiratory mortality rates, obscuring the trends in respiratory mortality related to pneumococcal vaccination.
*Methods.* We applied the inference method from our recent study of influenza-associated mortality in Russia to relate monthly respiratory mortality rates between 09/2010 and 08/2019 to monthly indices of influenza circulation and baseline rates of mortality not associated with influenza, allowing for a potential trend in the baseline rates starting 2015.
*Results.* Baseline rates of respiratory mortality not associated with influenza decreased with time starting from the 2014/15 season (with each season running from September to August), with the 2018/19 baseline rates of respiratory mortality per 100,000 individuals being lower by 12.41 (95% CI (10.6,14.2)) compared to the 2010-2014 period.

*Conclusions.* While improvement in care might have contributed to the reduction in the rates of respiratory mortality not associated with influenza in Russia, the above temporally consistent reduction is compatible with the gradual replacement of vaccine serotypes in the transmission and carriage of *S. pneumoniae* following the introduction of PCV13. Further work is needed to better understand the impact of PCV13 on the epidemiology of respiratory infections and related mortality in Russia.

**Introduction**

Pneumococcal conjugate vaccine (PCV13) was introduced in Russia in 2014, with high uptake in young children reported by 2016 [1]. Serotype replacement compared to carriage data from the pre-PCV13 years was found [2-4], with serotype replacement following pneumococcal vaccination also having taken place in other countries [5,6]. While rates of pneumonia following the introduction of PCV13 in Russia have not decreased, neither in children [1], nor overall [7], one can see a notable decline in the rates of mortality for respiratory causes in the years following the rollout of PCV13 [8]. Some of that decline may potentially be explained by a strong association between antibiotic, including multidrug resistance and PCV13 vaccine serotypes [3,4], with resistant infections with vaccine serotypes (compared to non-vaccine serotypes) being more likely to devolve into the most severe outcomes, including sepsis and death. Indeed, a major study of septic infections in St. Petersburg, Russia found that respiratory tract was the most common source of those infections, and *S. pneumoniae* was found very rarely for those cases [9].

In this study we aim to evaluate the trends in respiratory mortality rates in Russia following the introduction of PCV13 vaccination. Annual variability in influenza circulation may affect some of the changes in respiratory mortality rates, obscuring the above trends. For example, the rate of respiratory mortality in Russia during the 2014/15 season (September through August) was higher than during each of the four previous seasons (2010/11 through 2013/14), with that difference being related to the mortality associated with the major 2014/15 influenza season [10]. Thus, in order to study trends in respiratory (or pneumonia) mortality rates, one ought to adjust the observed rates for the effect of influenza circulation. Here, we apply the inference methodology from our recent study of influenza-associated mortality in Russia [10] to relate the monthly rates of respiratory mortality, provided by the Russian Federal State Statistics Service (Rosstat) [8] to the indices of monthly incidence of influenza A/H3N2, A/H1N1, and B in Russia (derived from the surveillance data from the Smorodintsev Research Institute of Influenza (RII) [11]), adjusting for baseline rates of respiratory mortality not associated with influenza. Moreover, we include terms for the trend in baseline rates of non-influenza associated respiratory mortality to examine the change in those rates following the introduction of PCV13. We also discuss the potential causes for the changes in respiratory mortality rates, including the effect of pneumococcal vaccination.

**Methods**

*Data*
Monthly data on mortality for respiratory causes in Russia were obtained from [8]. Monthly mortality counts for respiratory deaths were then converted to monthly rates of mortality per 100,000 individuals using population data from Rosstat (with annual population estimates interpolated linearly to estimate the population by month).

Weekly data on the rates of influenza/Acute Respiratory Illness (ARI), (грипп/ОРВИ) consultation per 10,000 individuals in Russia are available from [11]. Data on the weekly percent of respiratory specimens from symptomatic individuals that were RT-PCR positive for influenza A/H1N1, A/H3N2 and influenza B are also available from [11] (under the Laboratory Diagnostics link).

*Incidence proxies*

Only a fraction of individuals presenting with influenza/ARI symptoms are infected with influenza. We multiplied the weekly rates of influenza/ARI consultation per 10,000 individuals [11] by the weekly percentages of respiratory specimens from symptomatic individuals that were RT-PCR positive for each of influenza A/H1N1, A/H3N2 and B [11] to estimate the weekly incidence proxies for each of the corresponding influenza (sub)types:

*Weekly influenza (sub)type incidence proxy =*       (1)
*Rate of consultations for influenza/ARI * % All respiratory specimens that were RT-PCR positive for that influenza (sub)type*

As noted in [12], those proxies are expected to be *proportional* to the weekly population incidence for the each of the major influenza (sub)types (hence the name "proxy") – in fact, those proxies estimate the weekly rates of consultation for ARI associated with the corresponding influenza (sub)types, divided by the sensitivity of the RT-PCR test. Monthly incidence proxies for influenza A/H1N1, A/H3N2 and B were obtained as the weighted average of the weekly incidence proxies for those weeks that overlapped with a given month; specifically, for each influenza (sub)type and month, the incidence proxy for each week was multiplied by the number of days in that week that were part of the corresponding month (e.g. 7 if the week was entirely within that month), then the results were summed over the different weeks and divided by the number of days in the corresponding month. To relate the incidence proxies for the major influenza (sub)types to monthly mortality rates, we first shift the weekly incidence proxies by one week forward to accommodate for the delay between infection and death [12], then use the shifted weekly incidence proxies to obtain the corresponding monthly incidence proxies as above.

The relation between an incidence proxy and the associated mortality may change over time. In particular, influenza B is characterized by the circulation of *B/Yamagata* and *B/Victoria* viruses. It is known that the age distribution for the *B/Yamagata* infections is notably older than for the *B/Victoria* infections [13,14]. Correspondingly, the relation between the incidence proxy (which reflects influenza incidence in the general population) and influenza-related mortality (which for influenza B largely reflects mortality in older individuals) may be quite different for influenza *B/Yamagata* compared to influenza *B/Victoria*. While there are no data on the weekly percentages of *B/Yamagata* and *B/Victoria* among the tested respiratory specimens in [11], such whole-season data are available in [11]. For each influenza season (running from September to June), we obtain the proportions of influenza B specimens from that season that were for *B/Yamagata* and *B/Victoria* (using data from the last reported week during that season [11]), and multiply the weekly incidence proxy for influenza B during that season by the corresponding proportions to estimate the weekly incidence proxy for each of influenza *B/Yamagata* and *B/Victoria*. Finally, the 2014/15 season was characterized by the global circulation of a novel A/H3N2 variant. Mortality for that variant is potentially different from the mortality for the previously circulating A/H3N2 strains. Correspondingly, to relate A/H3N2 to respiratory mortality, we split the A/H3N2 incidence proxy into two: one (called $A/H3N2^1$) equaling the A/H3N2 proxy between 09/2010 through 08/2014, zero for subsequent months; the other (called $A/H3N2^2$) equaling the A/H3N2 proxy between 09/2014 through 08/2019, zero for previous months. Figure 1 plots the monthly incidence proxies for influenza A/H3N2 (two proxies), A/H1N1, *B/Yamagata* and *B/Victoria* between 09/2010 and 08/2019 (108 months).

***Inference Model***

Let $M(t)$ be the *average daily* respiratory mortality rate per 100,000 during month $t$ (with $t = 1$ for 09/2010, $t = 108$ for 08/2019), and $A/H3N2^1(t), A/H3N2^2(t), A/H1N1(t), B/Victoria(t), B/Yamagata(t)$ be the incidence proxies for the different influenza (sub)types on month $t$ as described in the previous subsection. The inference model in [10,12] suggests that

$$M(t) = \beta_0 + \beta_1 \cdot A/H3N2^1(t) + \beta_2 \cdot A/H3N2^2(t) + \beta_3 \cdot A/H1N1(t) + \beta_4 \cdot B/Victoria(t) + \beta_5 \cdot B/Yamagata(t) + Baseline + Noise \quad (2)$$

Here the noise is white noise (linear regression), and $Baseline$ is the baseline average daily rate of respiratory mortality per 100,000 not associated with influenza circulation. We assume that this rate is *periodic* with yearly periodicity, excpet for the potential trend starting 2015. We will model it as

$$Baseline(t) = \beta_6 \cdot \cos\left(\frac{2\pi t}{12}\right) + \beta_7 \cdot \sin\left(\frac{2\pi t}{12}\right) + \beta_8 \cdot \text{Jan}(t) + SE(t) + Trend \quad (3)$$

Here *Jan* is a variable equaling 1 for the month of January, 0 otherwise. The reason for including this variable is that the monthly (rather than annual) mortality data in [8] is operational, with some of the mortality not registered during a given calendar year being added to January of the next year [15]. The (temporal) trend is modeled as a quadratic polynomial in the month starting 01/2015 (thus the month for the trend equals 0 prior to 01/2015, it equals 1 for 01/2015, it equals 13 for 01/2016 etc., and the trend is a quadratic function of that month). Finally, the summer effect $SE(t)$ equals 1 for the month of July, 2 for the month of August, and 0 for other months. The reason for including this variable is that there is a consistent dip in respiratory mortality (Figure 2) during the months of July and August (particularly August), presumably having to do with the decline in the transmission of respiratory viruses when schools are closed/weather is hot, and this dip cannot be accommodated by the trigonometric model in eq. 3. While this variable wasn't included in the model in [10], its inclusion results in a significant improvement in the model fit, and excluding this variable has a very minor effect on the estimation of the trend in respiratory mortality following the introduction of PCV13.

## Results

Figure 1 plots the monthly proxies for the incidence of influenza A/H3N2 (split into two as described in Methods), A/H1N1, *B/Yamagata*, and *B/Victoria* during our study period (09/2010 through 08/2019).

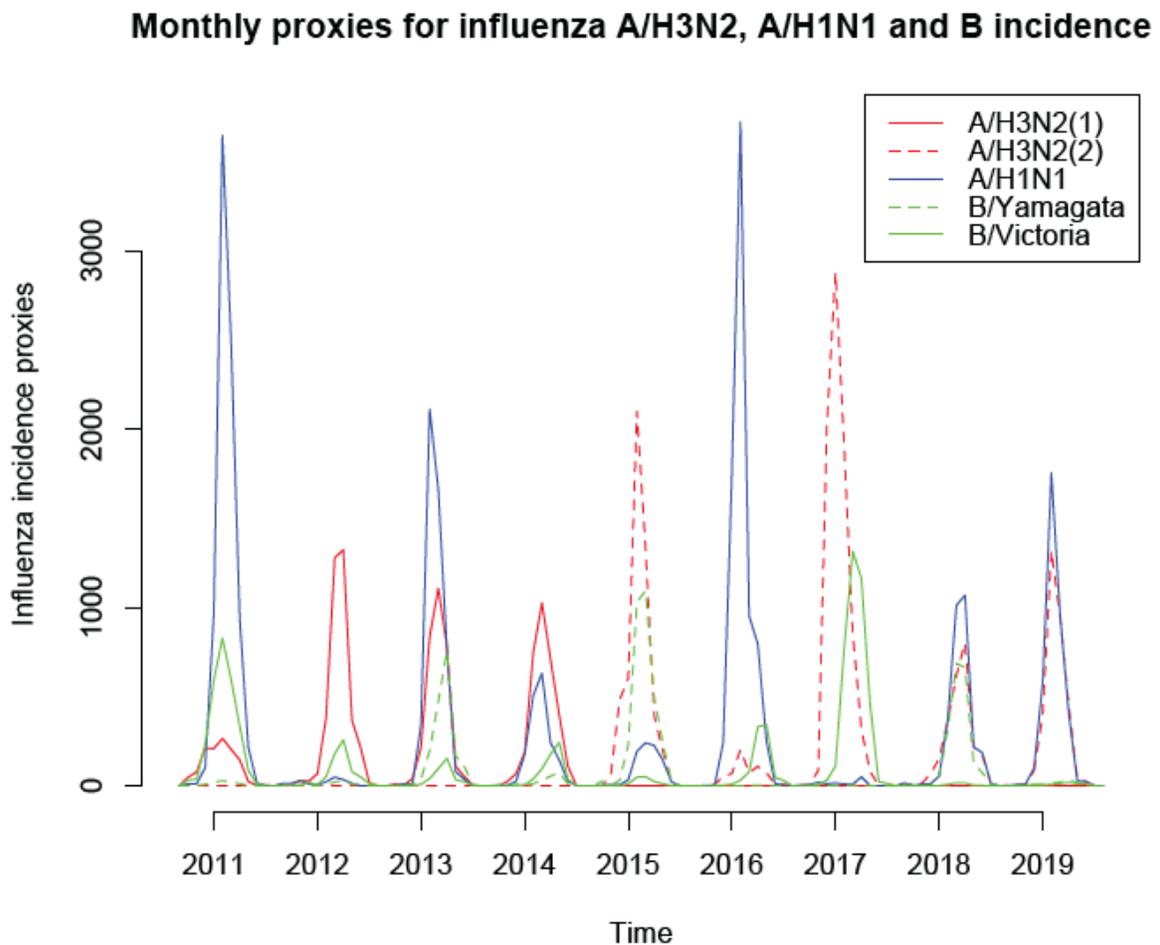

**Figure 1:** Monthly proxies for the incidence of influenza A/H3N2 (split into the 09/2010 though the 08/2104 period and the 09/2014 through the 08/2019 period), A/H1N1, *B/Yamagata*, and *B/Victoria* between 09/2010 through 08/2019.

Figure 2 presents the fits for the model in eq. 2 for the *average daily* rates of respiratory mortality per 100,000 individuals by month (black curve) for the months of 09/2010 through 08/2019. Those model fits were largely temporally consistent save for the early part of the study period, which may partly have to do with data quality for that period --- see Discussion. Figure 2 also exhibits the average daily baseline rates (by month) of respiratory mortality per 100,000 people not associated with influenza between 09/2010 though 08/2019. Those rates declined during the period following the introduction of PCV13 vaccination.

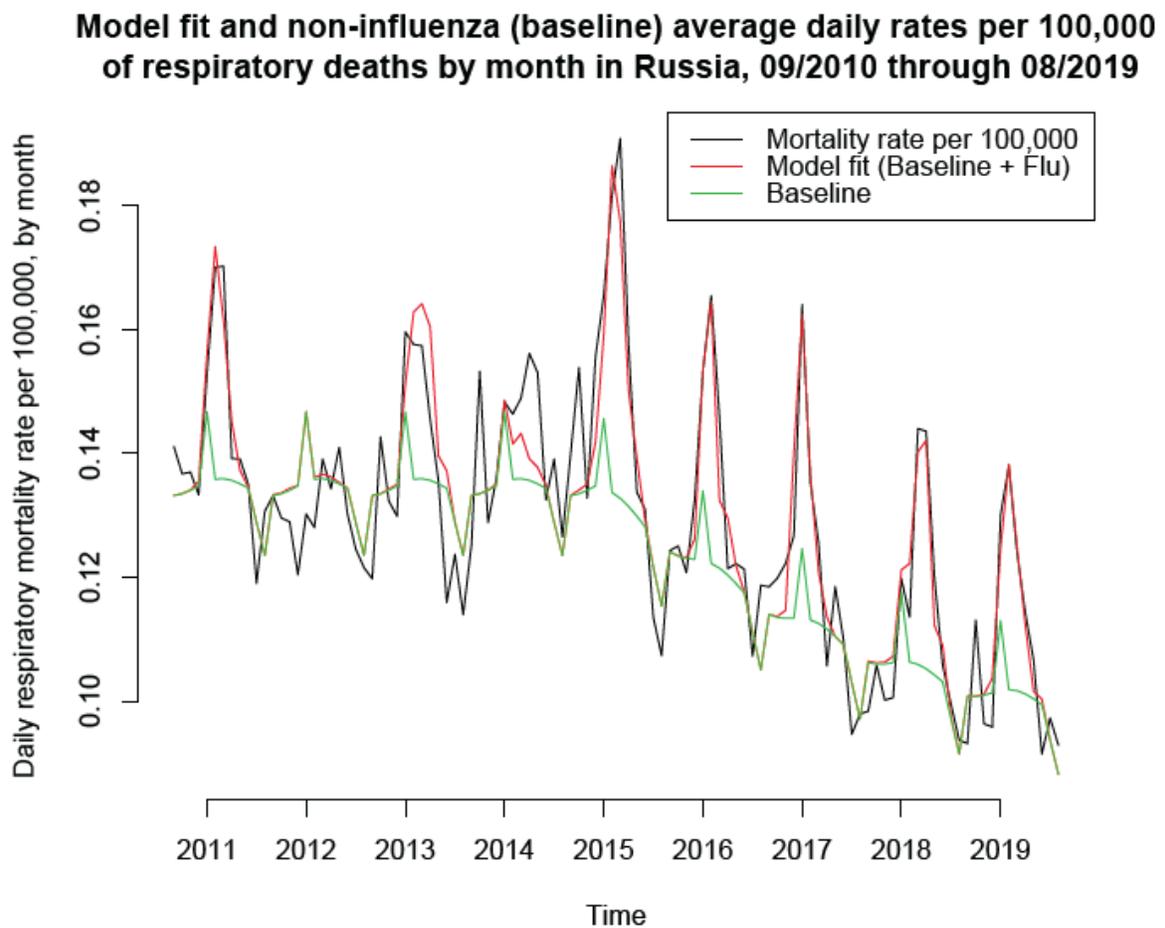

**Figure 2:** Average daily rates of respiratory mortality per 100,000 people by month (black curve); fits for the model in eq. 2 (red curve); average daily baseline rates (by month) of respiratory mortality per 100,000 people not associated with influenza (green curve) between 09/2010 though 08/2019.

Table 1 presents the estimates of the rates of influenza-associated respiratory mortality during each of the 2010/11 through the 2018/19 seasons (September through August for each season), as well as the expected seasonal baseline rates of respiratory mortality not associated with influenza. Those expected seasonal rates were modeled as constant for the 2010/11 through the 2013/14 seasons (save for a tiny change during the 2011/12 season due to the fact that 2012 was a leap year). Subsequently, those baseline rates declined, with a decline of 12.41 (10.6,14.2) respiratory deaths not related to influenza per 100,000 individuals during the 2018/19 season compared to the 2010/11 through the 2013/14 seasons.

| Season | Rate of influenza-associated respiratory mortality | Expected rate of non-influenza related respiratory mortality | Reduction in respiratory mortality rate following PCV-13 introduction |
|---|---|---|---|
| 2010/11 | 2.52 (1.5,3.6) | 49.01 (47.9,50.1) | |
| 2011/12 | 0.08 (-0.8,0.9) | 49.15 (48,50.2) | |
| 2012/13 | 2.76 (1.7,3.8) | 49.01 (47.9,50.1) | |
| 2013/14 | 0.66 (-0.1,1.4) | 49.01 (47.9,50.1) | |
| 2014/15 | 4.43 (3.5,5.4) | 47.87 (46.9,48.9) | 1.14 (0.8,1.5) |
| 2015/16 | 2.65 (1.9,3.3) | 44.05 (42.8,45.3) | 5.1 (3.9,6.3) |
| 2016/17 | 3.02 (1.6,4.4) | 40.63 (39.3,42) | 8.38 (6.8,9.9) |
| 2018/18 | 3.21 (2.6,3.8) | 38.19 (37,39.4) | 10.82 (9.3,12.3) |
| 2018/19 | 2.58 (2,3.1) | 36.6 (35,38.2) | 12.41 (10.6,14.2) |

**Table 1**: Seasonal rates of influenza-associated mortality in Russia for the 2010/11 through the 2018/19 seasons (September through August), expected seasonal rates of non-influenza associated respiratory mortality, and reduction in non-influenza associated mortality following the introduction of PCV13 starting the 2014/15 season.

**Discussion**

The pneumococcal conjugate vaccine PCV13 was introduced in Russia in 2014, with uptake increasing significantly during 2015 [1]. This was followed by a pronounced decline in respiratory mortality, including pediatric mortality [1], though not the rates of pneumonia hospitalization in Russia [1,7]. Some of the reasons for the above discrepancy may have to do with the fact that vaccine strains for PCV13 are much more drug-resistant, as well as multidrug-resistant compared to non-vaccine strains in Russia [3,4]. Influenza circulation affects the rates of respiratory mortality, as well as changes in those rates with time. To better understand the trends in respiratory mortality not associated with influenza circulation in Russia, we applied the inference model in [10,12] to estimate the (baseline) rates of non-influenza respiratory mortality, as well as trends in those baseline rates. We found a consistent decline in the rates of non-influenza respiratory mortality starting 2015. Moreover, this finding was robust with regard to several assumptions made in our inference model.

A key question is related to the causes behind the aforementioned decline in the rates of non-influenza respiratory mortality. Improvement in care could potentially contribute to decline in respiratory mortality rates. At the same time, changes in the epidemiology of pneumococcus, particularly reduction in the carriage of vaccine-type strains in children took place in Russia [2-4]. In the US, virtual disappearance of the winter holiday season bump in pneumonia mortality took place following the introduction of the PCV7 vaccine (Figure 2 in [12]), with that bump prior to the introduction of PCV7 ascribed to the transmission of pneumococcal strains from young children to their grandparents during the holiday season. In Russia, transmission of *S. pneumoniae* from young children to older individuals might be even greater than in the US due to differences in contacts between those age groups for the two countries. Additionally, non-vaccine strains of *S. pneumonia* in children in Russia are much less antibiotic-resistant compared to vaccine strains for PCV13

[3,4]. All of this suggests that the decline in non-influenza respiratory mortality rates estimated in this study is also consistent with the replacement of vaccine serotypes in the transmission and carriage of *S. pneumonia*, with that replacement being more pronounced year-to-year as more as more and more young children are protected by the vaccine [5]. Further work is needed to better understand the impact of PCV13 on the epidemiology of *S. pneumonia* (including serotype replacement in the elderly population) and related mortality in Russia.

Our paper has some limitations. We only had access to monthly mortality data; moreover, those data are operational, with some delays in reporting, and some unreported deaths during a given calendar year being reported for January of the next year [15]. Moreover those data were made available after the start of the study period [15], with the earlier data extracted retrospectively. Influenza surveillance data can also be subject to noise such as the discrepancy between the timing of specimen collection and testing/reporting. For the 2013/14 season, influenza circulation was still significant by week 20 of 2014, with 18% of respiratory specimens testing positive for influenza during that week [11]; however, no surveillance data for the subsequent weeks during that season are available in [11]. All of this might explain some lack of temporal consistency in the model fit (Figure 2), particularly during the early part of the study period when the quality of the data may be more questionable. Finer mortality data stratified by week/age group are needed to get a more comprehensive understanding of decline in the rates of respiratory mortality no associated with influenza in the period following the introduction of PCV13 in Russia.

We believe that despite the above limitations, our results suggest a robust decline the rates of respiratory mortality not associated with influenza following the introduction of PCV13 in Russia. That decline is consistent with replacement of vaccine serotypes of *S. pneumoniae* that are more drug-resistant compared to non-vaccine serotypes in Russia. Further work is needed to better understand the impact of PCV13 on the epidemiology of *S. pneumonia* and related severe outcomes, including mortality in Russia.